# Towards a graphene-based quantum impedance standard

C.-C. Kalmbach[1], J. Schurr[1], F. J. Ahlers[1], A. Müller[1], S. Novikov[2],
N. Lebedeva[2], and A. Satrapinski[3]

[1]Physikalisch-Technische Bundesanstalt, Bundesallee 100, D-38116 Braunschweig, Germany
[2]Department of Micro- and Nanosciences, Aalto University, Micronova, Tietotie 3, 02150 Espoo, Finland
[3]MIKES, Tekniikantie 1, P.O.Box, 02151 Espoo, Finland

Precision measurements of the quantum Hall resistance with alternating current (ac) in the kHz range were performed on epitaxial graphene in order to assess its suitability as a quantum standard of impedance. The quantum Hall plateaus measured with alternating current were found to be flat within one part in $10^7$. This is much better than for plain GaAs quantum Hall devices and shows that the magnetic-flux-dependent capacitive ac losses of the graphene device are less critical. The observed frequency dependence of about $-8\times10^{-8}$/kHz is comparable in absolute value to the positive frequency dependence of plain GaAs devices, but the negative sign is attributed to stray capacitances which we believe can be minimized by a careful design of the graphene device. Further improvements thus may lead to a simpler and more user-friendly quantum standard for both resistance and impedance.

Graphene is probably the most fascinating electronic material discovered in the last decades [1-3]. Among its various unique properties, an anomalous 'half-integer' quantum Hall effect (QHE) is most interesting for metrology, where the fact that the Hall resistance is quantized and depends only on fundamental constants is utilized for the representation and maintenance of the resistance unit, the ohm. Typically, two-dimensional electron systems (2DES) realized in GaAs/AlGaAs heterostructures [4] are used for this purpose. The required relative measurement uncertainty of better than 1 part in $10^8$ is, however, only obtained at strong magnetic fields around 10 tesla and at temperatures of 1.4 kelvin and below. In contrast, in graphene the cyclotron energy splitting between the Landau levels (which is the main factor determining the robustness of the quantized Hall resistance (QHR)) is so large that fingerprints of the QHE are even observed at room temperature [5]. Thus, with graphene a highly precise QHR standard working at low magnetic fields and temperatures above 4 kelvin is conceivable, which would be an enormous advantage for practical metrology. In fact, when measuring with direct current (dc), it has been demonstrated already that the precision of the QHE in high quality graphene devices matches that of GaAs devices [6-9]. However, in the forthcoming fundamental constant-based redefinition of the Système International d'Unités (SI) [10], also the impedance units (capacitance and inductance) will be traced to fundamental constants [11]. The most direct way to represent the impedance units is to use a quantum Hall resistance measured with alternating current (ac QHR). This has two advantages. Firstly, deriving the resistance and impedance units from the same quantum effect improves the consistency of the SI. And secondly, using the same QHE device at dc and at ac in one and the same cryomagnetic system would constitute a practical and economical advantage. Therefore, the question naturally arises whether graphene can replace GaAs also in the realm of impedance units, leading to an at least equally precise, but more user-friendly and widely usable quantum impedance standard, applicable even in industry and calibration service labs.

In this paper, we demonstrate precision ac measurements of the QHE in graphene. The precision achieved cannot be taken for granted as experience with early ac QHR measurements on GaAs devices has shown [12]. While there is no theoretical evidence or prediction that the quantized Hall resistance should exhibit significant inherent frequency dependence in the range of a few kHz, the capacitive coupling of the 2DES to the unavoidable metallic environment as well as within the device itself can limit measurement uncertainty. Strictly speaking, it is the dissipation factor of parasitic capacitances which can lead to frequency-dependent deformations of the QHR plateaus and to deviations from the quantized dc resistance value. To eliminate the influence of these parasitic capacitances in the case of GaAs devices, a double-shielding method and an alternative extrapolation method had been developed to achieve an uncertainty comparable to dc resistance calibrations [13,14]. Both methods are elaborate and therefore not widely used. Our first results obtained with graphene devices are unexpectedly good and in fact even better than those of plain (i.e. not specially shielded) GaAs devices. They demonstrate that it is indeed promising to develop a graphene-based impedance standard which is much less affected by capacitive effects and can possibly be applied without the elaborate methods required for GaAs, and at the same time can be operated under the same relaxed temperature and magnetic field requirements as are envisioned for the application of graphene as a dc resistance standard [6-9].

The measurements presented in this study were carried out on a large-area Hall bar device (800 µm x 200 µm) lithographically patterned on a graphene film grown on the silicon-terminated face of a 4H silicon carbide substrate [15]. The film had been grown in argon at atmospheric pressure, at a temperature of 1650 °C within 5 minutes. Its thickness and quality were assessed by Auger spectroscopy, utilizing the Si to C peak ratio [16] to confirm the presence of single layer graphene. According to these measurements the total coverage of the SiC surface by graphene was a bit less than one which means that the growth process was completed before the second graphene layer started to grow. From Raman spectra, a 2D-peak width of 40 cm$^{-1}$ also supports the presence of monolayer graphene [17]. Laser photolithography was utilized for the patterning of Hall bars and contacts. The graphene surrounding the Hall bars was completely removed by reactive ion etching in argon-oxygen plasma to minimize unwanted capacitive coupling between the Hall device and the surrounding areas of unused graphene. The direction of the Hall bars was aligned parallel to the terrace edges of the SiC substrate. Stable and low-resistance contacts were fabricated by a two-step Ti/Au (5/50 nm) metallization process using e-beam lithography and liftoff photolithography. Photochemical gating [18] was applied to tune the charge-carrier concentration by covering the sample with two polymers (first 300 nm PMMA resist and second 300 nm ZEP520A resist) and subsequent UV irradiation. More details of sample processing are described in [15]. To tune the carrier concentration to a state with optimal quantization at reasonably low magnetic flux densities, the sample underwent iterative cycles of UV illumination, dc resistance measurements at low temperatures, and (if illuminated for too long) thermal recycling at 170 °C (for 15 minutes).

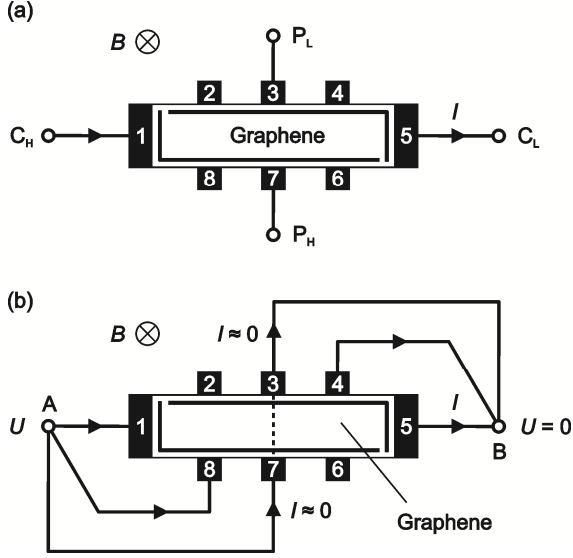

FIG. 1. (a) Scheme of the QHR device connected for the four-terminal dc measurement of $R_{xy}$. $R_{xx}$ is measured at contacts 8 and 7, respectively. (b) Triple-series connection scheme for ac measurement of the QHR.

All measurements were carried out with the device in liquid $^3$He of bath temperature $T = 0.7$ K [19] in a cryostat system equipped with a superconducting solenoid and coaxial measuring leads. A schematic drawing of the contacted device is shown in Fig. 1a. Four-terminal dc measurements of the Hall resistance $R_{xy}$ (at contact pair 7, 3) and the longitudinal resistance $R_{xx}$ (at contact pair 8, 7) were performed by a 6½-digit scanning voltmeter while the source-drain current ($I_{dc} = 10$ µA between contacts 1, 5) was provided by a battery-operated current source. A well-pronounced plateau for filling factor $v = 2$ and a vanishing longitudinal resistance are observed at a magnetic flux density above $B \approx 8$ T (Fig. 2), providing the most direct evidence [1] that the device indeed consists of monolayer graphene. From the slope of the Hall resistance at low magnetic flux densities (see the dashed line in Fig. 2), the electron concentration of the device was determined as $n = 6.3 \times 10^{11}$ cm$^{-2}$, predicting $v = 2$ at around $B = 13$ T (disregarding the fact that the $v = 2$ state continues to much higher magnetic flux densities due to the Fermi level pinning effect reported in [20]). The electron mobility of $\mu = 1730$ cm$^2$/Vs was determined from $n$ and $R_{xx}$ at zero magnetic field. A precision dc measurement of $R_{xy}$ with a Cryogenic Current Comparator bridge (CCC) was performed at $B = 13.5$ T, applying a current of $I_{dc} = 30$ µA. The Hall resistance was measured against a 100 Ω reference resistor which in turn had been calibrated against a GaAs quantum Hall device. The measurement revealed an excellent quantization of the $v = 2$ plateau within an uncertainty of 7 parts in $10^9$ (coverage factor $k = 1$), revealing the very good dc characteristics of this graphene device.

For ac measurements, an in-house developed four-terminal-pair coaxial ac resistance bridge whose design is described elsewhere [21] was used to compare the quantum Hall resistance of the graphene device with a 12.9 kΩ reference resistor which, in turn, had been measured at ac against well-characterized double-shielded GaAs QHR devices. Here, the graphene device was connected according to the triple-series connection scheme [22,23] (Fig. 1b), in which the respective equipotential terminals (contacts 1, 7, 8 and contacts 3, 4, 5) are connected outside of the cryostat to the nodes A and B, respectively. Such a connection scheme is standard for precision ac measurements of the QHR, since, due to the properties of the QHE, the current in the middle potential leads is practically zero, similar to a four-terminal-pair measurement of a conventional resistor. A four-terminal-pair ac measurement of the QHR (with connections as in Fig. 1a) would suffer a considerable error because the impedance between associated current and potential terminals is not low but equal to the Hall resistance [21]. Note that for ac measurements bond wires of unused contacts (contacts 2 and 6 in our case) have to be removed, because otherwise the open-circuited lead capacitances would draw considerable ac currents through the device. These currents do not contribute to the current measurement, but cause an additional Hall voltage which would lead to a wrong Hall resistance [21].

Figure 3a shows precision ac measurements of the QHR at the $v = 2$ plateau of the graphene device, presented as the relative deviation $\Delta r = (R_{xy} - R_K/2)/(R_K/2)$ from $R_K/2 = h/2e^2$. A typical ac QHR plateau of a GaAs device is shown for comparison in Fig. 3b. Note that no double-shielding method is applied in these measurements and that only the GaAs device had a metal backplane (which was standard before the double-shielding and extrapolation methods were developed). Several remarkable differences are striking. First of all, the graphene plateau is flat within 1 part in $10^7$, or even better, at all measured frequencies over a range of several tesla, exhibiting no frequency-dependent curvature. In contrast, GaAs devices show a strong plateau curvature proportional to frequency [24] which originates from the dissipative part of the capacitances between the 2DES and surrounding metals [25]. In GaAs devices, this effect compromises the precision of the application as an impedance standard unless it is eliminated by the double-shielding method [13]. In the graphene device, however, the magnetic-flux-dependent capacitive dissipation is very small even without any counteraction. This could be due to the smaller Hall bar size (800 µm × 200 µm, compared to 2600 µm × 800 µm for GaAs), the absence of a metallic backplane on the chip carrier (present in the GaAs device), and/or a smaller dissipation factor of the capacitance between the 2DES and surrounding metals. A second remarkable difference is that the graphene plateau is much wider than in GaAs devices. This was first observed by Janssen et al. [20] in dc QHR measurements and is explained by charge transfer between the substrate and the epitaxial graphene pinning the filling factor over a wide range of magnetic flux density.

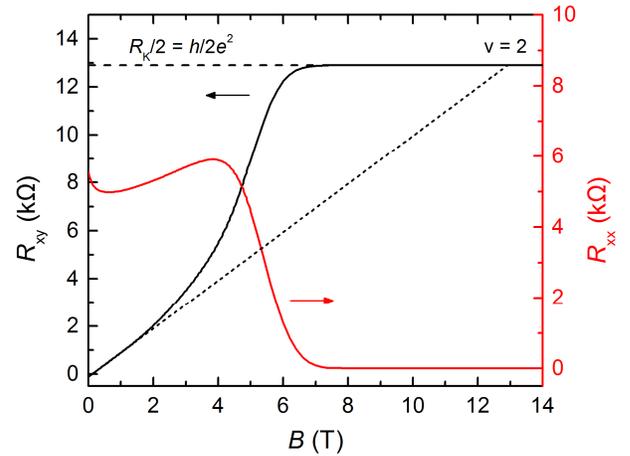

FIG. 2. Hall resistance (left-hand scale) and longitudinal resistance (right-hand scale) of a SiC-graphene device measured at a bath temperature of $T = 0.7$ K and a direct current of $I_{dc} = 10$ µA. The dashed line indicates the slope of the Hall resistance at low magnetic flux densities.

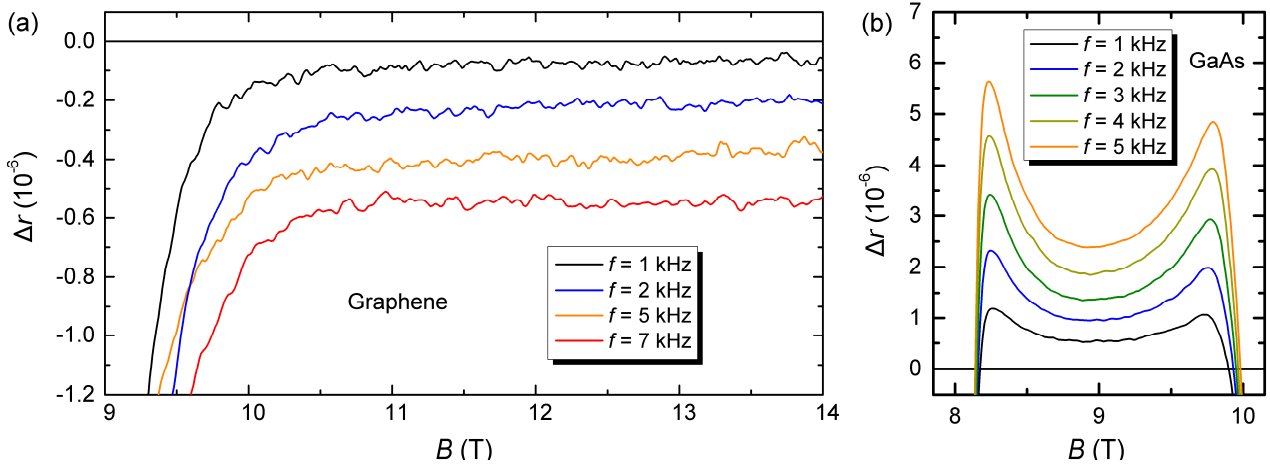

FIG. 3. (a) Relative difference of the quantum Hall resistance at the graphene plateau $\nu = 2$ from the quantized dc value, measured as a function of magnetic flux density at different frequencies as indicated, at a bath temperature of $T = 0.8$ K and a current of $I = 10$ µA rms. The noise of the measurement amounts to $1.2\times10^{-8}$ and is mainly due to the thermal noise of the room-temperature reference resistor. (b) The corresponding measurement at a GaAs device with a metal backplane at a bath temperature of 0.3 K and a three times longer integration time. Note the different scales of the ordinates.

Furthermore, and similar to GaAs devices not double shielded, the quantized Hall resistance of the graphene device exhibits a linear frequency dependence, as presented in Fig. 4. The superimposed resonance at $f = 3.3$ kHz is an artifact which we attribute to bond wire resonances. Such resonances are particularly seen in each ac contact resistance measurement and their amplitudes converge to zero with decreasing magnetic flux density. This clearly shows that the resonances are caused by current-driven electro-mechanical vibrations of bond wires in a magnetic field[1] [26,27], and they are not related to graphene. The bond wires are 5-6 mm long due to the small size of the Hall bar compared to the large chip and a carrier dedicated to dc measurements. This is already too long and future graphene devices for ac measurements will consider this aspect adequately. In the following, the data in the affected frequency range are not taken into account, and the uncertainty is increased appropriately.

In general, a linear frequency dependence of the QHR is attributed to the dissipation factor of parasitic capacitances [12, 25]. For the application as a quantum impedance standard, negligible frequency dependence is desired. Our graphene device shows a frequency coefficient of about $-8\times10^{-8}$/kHz (Fig. 4). The absolute value of this frequency coefficient is even lower than that of GaAs devices not double shielded and with a metal backplane ($+50\times10^{-8}$/kHz, see Fig. 3) [24] and comparable to that of GaAs devices not double shielded without a metal backplane. Furthermore, the frequency coefficient of the graphene device is negative whereas all previously reported GaAs devices show a positive frequency coefficient [24, 12]. In fact, parasitic capacitances can lead to both positive and negative frequency dependence (while the dissipation factor is always positive). Negative frequency dependence is attributed to dissipative capacitances between contact pads or bond wires at opposite equipotential sides of the QHR device (i.e. capacitances *in parallel* to the Hall resistance). A positive contribution to the frequency dependence is attributed to dissipative capacitances between the 2DES and surrounding metals (i.e. capacitances *in series* to the Hall resistance) [25]. We expect the positive contributions to decrease with decreasing size of the QHR device, whereas the negative contributions should increase. Although the graphene device used here is relatively large for its kind, it is much smaller than conventional GaAs devices. Consequently, the negative contributions of our graphene device dominate.

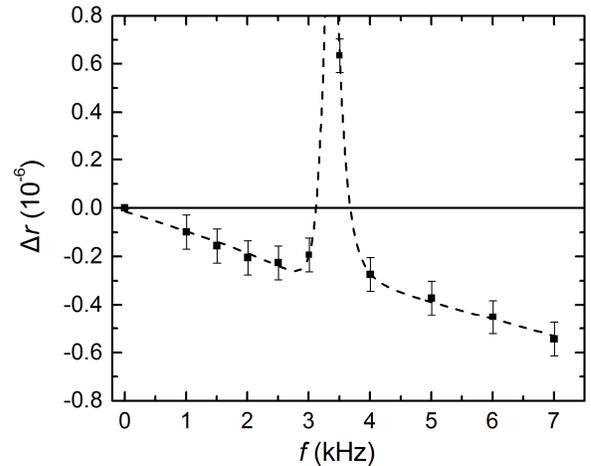

FIG. 4. Frequency dependence of the quantized Hall resistance at $B = 13.5$ T (corresponding to $\nu = 2$), at a bath temperature of $T = 0.8$ K and a current of $I = 10$ µA rms, measured as a function of frequency. The dashed line is a least-squares fit of a model function. The uncertainty bars are ten times larger than the measurement uncertainty in order to account for the incomplete control of bond wire vibrations.

---

[1] The common solution of using two parallel bond wires of different length is not applicable here because, due to their small distance, the bond wires are coupled by their mutual inductance and still vibrate.

More detailed studies of the different contributions and of the dissipation factor of the SiC substrate and the ZEP520 resist in comparison to a GaAs substrate will be carried out in the future. Measurements of the magnetocapacitance of the 2DES and the associated dissipation factor [28] will enable further insight into the ac properties of graphene devices. Careful engineering of device and contact dimensions may allow achieving a frequency-independent quantum Hall resistance without need for a complex shielding method and the respective adjustments, leading to a graphene-based impedance standard which is as accurate as double-shielded GaAs devices, but more user-friendly and simpler to operate.

In summary, we have shown that already the first ac measurements of the quantum Hall resistance in graphene at filling factor $\nu = 2$ exhibit a wide plateau which is remarkably flat in the kHz frequency range and particularly exhibits no frequency-dependent curvature. It shows only a weak frequency dependence, comparable to that of plain GaAs devices without any complex shielding methods. We therefore expect that suitably optimized graphene devices should not only outperform the conventional devices in terms of lower magnetic field and higher temperature operation but also with respect to a simpler and more user-friendly operation with alternating current. A fundamental-constant-based quantum standard for both resistance and impedance would thus become reality, making calculable impedance artifacts dispensable and supporting the forthcoming redefinition of the SI.

This research has been performed within the EMRP project SIB51, *GraphOhm*. The EMRP is jointly funded by the EMRP participating countries within EURAMET and the European Union. We also gratefully acknowledge the help with cryostat technology by V. Bürkel, and the help with dc precision measurements by B. Schumacher and E. Pesel.